\documentstyle[prd,aps]{revtex}
\begin{document}
\preprint{\vbox{\hbox{LPTENS-97/??}    
\vspace{-.4cm}
                \hbox{CTP-TAMU-??/97}  \vspace{-.4cm}
                \hbox{IP-BBSR/97/??}   \vspace{-.4cm}
                \hbox {hep-th/9707182}}}
\twocolumn[\hsize\textwidth\columnwidth\hsize\csname
@twocolumnfalse\endcsname

\title{\noindent{\small \hfill\hfill CTP TAMU-31/97}
\newline
\noindent{\small \phantom{a}  \hfill CPTH-S547.0797}
\newline
\noindent{\small \phantom{a} \hfill \hfill SISSARef. 94/97/EP}
\newline
\noindent{\small \phantom{a} \hfill \hfill hep-th/9707182}
\bigskip\newline
Cosmological Solutions, p-branes and the Wheeler DeWitt Equation}

%\title{Cosmological Solutions, p-branes and the Wheeler-DeWitt Equation}

\author{H. L\"u$^{(1)}$, J. Maharana$^{(2,3)}$, 
S. Mukherji$^{(4)}$ and C. N. Pope$^{(4,5)}$}

\address{$^{(1)}$ {\it Laboratoire de Physique Th\'eorique de l'\'Ecole
Normale Sup\'erieure, Paris CEDEX 05.}}

\address{$^{(2)}$ {\it Institute of Physics, Bhubaneswar, India}}

\address{$^{(3)}$ {\it Centre Physique Theorique, Ecole Polytechnique, 
91128 Palaiseau, France}}

\address{$^{(4)}$ {\it Physics Department, Texas A \& M University, 
College Station, USA. }} 

\address{$^{(5)}$ {\it SISSA, Via Beirut No. 2-4, 34013 Trieste,
Italy}}

%\date{\today}
\maketitle

\begin{abstract}
     The low energy effective actions which arise from string theory
or M-theory are considered in the cosmological context, where the
graviton, dilaton and antisymmetric tensor field strengths depend only
on time.  We show that previous results can be extended to include
cosmological solutions that are related to the $E_N$ Toda equations.
The solutions of the Wheeler-DeWitt equation in minisuperspace are
obtained for some of the simpler cosmological models by introducing
intertwining operators that generate canonical transformations which
map the theories into free theories.  We study the cosmological
properties of these solutions, and also briefly discuss
generalised Brans-Dicke models in our framework. The cosmological
models are closely related to $p$-brane solitons, which we discuss in
the context of the $E_N$ Toda equations.  We give the explicit
solutions for extremal multi-charge $(D-3)$-branes in the truncated
system described by the $D_4 =O(4,4)$ Toda equations.

\end{abstract}\vskip2pc]
%*\documentstyle[11pt]{article}
%%%\documentstyle[11pt,epsf]{article}

%%%%% change page size and line spacing %%%%
%*\textwidth=6in
%*\hoffset=-.55in
%*\textheight=9in
%*\voffset=-.8in
%*\def\baselinestretch{1.4}
%%%%%%%%%%%%%%%%%%%%%%%%%%%%%%%%%%%%%%%%%%%%

%%%%% number equations by section %%%%%%%%
%\makeatletter
%\@addtoreset{equation}{section}
%\makeatother
%\renewcommand{\theequation}{\thesection.\arabic{equation}}
%%%%%%%%%%%%%%%%%%%%%%%%%%%%%%%%%%%%%%%%%%%

%*\def\dalemb#1#2{{\vbox{\hrule height .#2pt
%*        \hbox{\vrule width.#2pt height#1pt \kern#1pt
%*                \vrule width.#2pt}
%*        \hrule height.#2pt}}}
%*\def\square{\mathord{\dalemb{6.8}{7}\hbox{\hskip1pt}}}

\def\half{{\textstyle{1\over2}}}
\def\a{\alpha} 
\def\b{\beta} 
\def\g{\gamma} 
\def\d{\delta} 
\def\e{\epsilon}
\def\z{\zeta} 
\def\h{\eta} 
\def\q{\theta} 
\def\i{\iota} 
\def\k{\kappa}
\def\l{\lambda} 
\def\m{\mu} 
\def\n{\nu} 
\def\x{\xi} 
\def\p{\pi} 
\def\r{\rho}
\def\s{\sigma} 
\def\t{\tau} 
\def\u{\upsilon} 
\def\f{\phi} 
\def\c{\chi} 
\def\y{\psi}
\def\w{\omega}      
\def\G{\Gamma} 
\def\D{\Delta} 
\def\Q{\Theta} 
\def\L{\Lambda}
\def\X{\Xi} 
\def\P{\Pi} 
\def\S{\Sigma} 
\def\U{\Upsilon} 
\def\F{\Phi} 
\def\Y{\Psi}
\def\C{\Chi} 
\def\W{\Omega}     
\def\la{\label} 
\def\ci{\cite} 
\def\re{\ref}
\def\se{\section} 
\def\sse{\subsection} 
\def\ssse{\subsubsection} 
\def\nn{\nonumber} \def\bd{\begin{document}} \def\ed{\end{document}}
\def\ds{\documentstyle} 
\def\fr{\frac} \def\bl{\bigl} \def\br{\bigr}
\def\Br{\Bigr} \def\Bl{\Bigl} 
\def\bm{\bibitem}
\def\na{\nabla}
\def\pa{\partial} \def\ov{\overline}
\def\ie{{\it i.e.\ }} 
\newcommand{\be}{\begin{equation}} 
\newcommand{\ee}{\end{equation}} 
\def\ba{\begin{array}}
\def\ea{\end{array}}
\def\ft#1#2{{\textstyle{{\scriptstyle #1}\over {\scriptstyle #2}}}}
\def\fft#1#2{{#1 \over #2}}
\def\del{\partial}
\def\sst#1{{\scriptscriptstyle #1}}
\def\oneone{\rlap 1\mkern4mu{\rm l}}
\def\e7{E_{7(+7)}}
\def\td{\tilde}
\def\wtd{\widetilde}\def\R{\rlap{\rm I}\mkern3mu{\rm R}}
\def\im{{\rm i}}
\def\bog{Bogomol'nyi\ }

\newcommand{\ho}[1]{$\, ^{#1}$}
\newcommand{\hoch}[1]{$\, ^{#1}$}
\newcommand{\bea}{\begin{eqnarray}} 
\newcommand{\eea}{\end{eqnarray}} 
\newcommand{\ra}{\rightarrow}
\newcommand{\lra}{\longrightarrow}
\newcommand{\Lra}{\Leftrightarrow}
\newcommand{\alp}{\alpha^\prime}
\newcommand{\bp}{\tilde \beta^\prime}
\newcommand{\tr}{{\rm tr} }
\newcommand{\Tr}{{\rm Tr} } 
\newcommand{\NP}{Nucl. Phys. }
\newcommand{\tamphys}{\it Center for Theoretical Physics,
Texas A\&M University, College Station, Texas 77843}

\section{Introduction}
%{\em I. Introduction.}\hspace{0.5cm}

   There has been considerable attention given to the investigation of
the cosmological consequences of string theory. It is hoped that
string theory will provide answers to deep questions in quantum
gravity and therefore, it is natural that the problem of the evolution
of the Universe at early epochs be addressed in the string theory
framework [1-4]. One of the important and intriguing problems for
cosmology is to explain the mechanism of inflation. When one tries to
understand inflation from the perspective of string theory, it is
hoped that it should arise naturally from the theory itself. It has
been recognised that the dilaton field might play an important r\^ole
in the explanation of inflation. However, the dilaton also determines
the coupling constant in string theory, and therefore it must decouple
at late times so that the well-known results of late-time cosmology
are not affected by dilaton interactions, in view of the fact that it
can affect masses and coupling constants and other parameters at late
times.  This has motivated a search for mechanisms that can account
for the dilaton decoupling.

    A mechanism has been proposed in the pre-big-bang
scenario \cite{mggv} which exploits symmetries that are particular to
string theory. The starting point is the tree level string effective
action for the dilaton and graviton.  There exist solutions that
describe an expanding Universe with decceleration.  This solution can
be related, by means of stringy symmetries, scale factor duality
and a time reversal transformation, to a solution which describes a
Universe that is expanding and accelerating. An attractive scenario
emerges from these two symmetry-related solutions in which the
Universe begins with rapid expansion, {\it i.e.}\ there is a pole
driven inflation, with a deccelarating expansion at later times,
eventually connecting smoothly to an FRW Universe.

Recently, some attention has been focussed on the investigation of the
cosmological aspects of the $p$-branes that arise as classical
solutions of string effective actions or those of supergravity
theories [6-14]. It is found that these solutions can be classified into two
broad catagories, depending on whether the solution is supported by
field strengths carrying electric charges or magnetic charges.  In
some cases, dualities can relate the two kinds of solution.  Although
the field equations appear to be quite complicated, even in the
cosmological context where the fields depend only on time,
nevertheless wide classes of exact classical solutions can be
obtained.  In fact the equations of motion can be cast into the form
of one-dimensional Liouville or Toda equations [9-11].  In particular,
in certain cases one encounters the $SL(N+1,\R)$ Toda equations
\cite{lpsln,lmpw}.  Later, we shall show that this can be extended to
the $E_N$ Toda equations.

      Since one would like to understand the evolution of the Universe
at very early times, it is natural to consider string cosmology in a
quantum framework.  One approach is to solve the Wheeler-DeWitt (WDW)
equation in a minisuperspace, and to examine the properties of the
solutions \cite{hl,gmv,mmp}.  One of the interesting applications of
quantum string cosmology is to provide a resolution for the graceful
exit problem, since no-go theorems have established that it cannot be
resolved in classical string cosmology in the pre-big-bang scenario
\cite{rbgv}.

The purpose of this article is to explore further the cosmological solutions
of string effective actions in the presence of generalised gauge potentials,
and to examine their properties.  First we review the classical
solutions and then we study them at the quantum level.   We shall show that
formal solutions to the WDW equation can derived in an elegant
manner in the minisuperspace model. We shall present explicit solutions in  
some simple cases as illustrative examples.
 
In perturbative string theory, there are global symmetries,
including the extensively-studied $O(d,d)$ T-duality symmetry.  This
symmetry makes it possible to solve the WDW equation in many cases,
allowing the wave function to be completely classified by the $O(d,d)$
quantum numbers \cite{kl}.  We shall analyse the WDW equation for a
wide class of cosmological models that arise in the low-energy
effective string theory.  The organising symmetry that naturally
appears in these cases is $SL(N+1, \R)$ or $E_N$; the $O(d,d)$
symmetry is not manifest here since we use Poincare duality to write
down different form fields that appear in the models. In what follows,
we shall discuss how the $SL(N+1, \R)$ symmetry helps us to solve the
WDW equation in minisuperspace.

The paper is organised as follows.  In section II, we discuss the
different cosmological models that arise naturally in string theories
or M-theory in their low energy limit.  In particular, in sections
IIA, IIB and IIC we review the single-charge, multi-charge and $SL(3,
\R)$ Toda models that have been analysed in detail in \cite{lmpw}. In
section IID, we present a new class of $E_N$ Toda models.  In section
II, we also introduce the notion of canonical transformations which
map sets of interacting classical equations into sets of free
equations. This, in turn, allows one to solve the equations of motion
in a very simple manner.  The quantum version of this canonical
transformation is introduced in section III with the help of an
intertwining operator closely following the work of \cite{anps}. This
operator maps an interacting quantum theory to a free theory through a
set of {\it non-unitary} transformations. This property of the
operator is then exploited to solve the WDW equations of some of the
models introduced in section II. We conclude the paper with a
discussion of Brans-Dicke theory extended to include form fields. In
particular, we show that solutions of the classical equations of
motion and also the solutions of the WDW equation can be obtained from
the models discussed in section II and III by simple rescalings and
redefinitions of fields.  In an appendix, we consider classes of $p$-brane
solutions that are closely related to the cosmological $E_N$ Toda models.
In particular, we show how to solve the $E_N$ Toda equations for extremal 
$(D-3)$-branes, taking a simplifying truncation to the $D_4=O(4,4)$ Toda
system as an explicit example that has not previously been presented in
the literature.

%%%%%%%%%%%%%%%%%%%%%%%%%%%%%%%%%%%%%%%%%%%%%%%%%%%%%%%%%%%

\section{Cosmological models with NS-NS or R-R form fields } 

The effective low-energy limits of string theory or M-theory compactified 
on tori give rise to maximal supergravities in lower dimensions.  
In \cite{lmpw,lmp}, many cosmological solutions were obtained and analysed. 
These included single-charge, multi-charge and $SL(N+1,\R)$ Toda models. 
It was found that the classical equations of motion could be reduced to 
a set of Liouville or Toda equations. 
In sections IIA, IIB and IIC we review these models and also, in section IID
we introduce a new class of models which are $E_N$ Toda
cosmological models. We also introduce a set of canonical transformations
which, at the classical level, mapss the Liouville or Toda theories
to theories governed by free Hamiltonians.

\subsection{Single-charge cosmological models}

The simplest cosmological model in $D$ dimension involves the 
metric, a dilaton and an $n$-rank field strength $F_n$ \cite{lmpw}. The 
action is given by 
\be 
S = \int d^Dx \sqrt{-g}[ R - {1\over 2}(\partial\phi )^2 - 
{1\over{2n!}}e^{a\phi}F_n^2] \ , 
\label{ac1}
\ee
where the constant $a$ can be parametrised as \cite{lpsol} 
\be
a^2 = \Delta - {2(n-1)(D-n-1)\over {D-2}}\ .
\label{avalue}
\ee
We will assume that all the fields depend only on time. The 
background metric is assumed to have the form
\be
ds^2 = -e^{2U}dt^2 + e^{2A}d{\bar s}^2 + e^{2B} dy^m dy^m \ ,
\label{metr}
\ee
where $d{\bar s}^2$ represents the $p$-dimensional metric on 
the spatial section of a $d$-dimensional spacetime, with $d = p+1$.
We shall consider spatial metrics of the maximally-symmetric form
\be
d{\bar s}^2 = {dr^2\over {1 -kr^2}} + r^2d\Omega^2 \ ,
\label{sec}
\ee
where $d\Omega^2$ is the metric on a unit $(p-1)$-sphere. Without 
loss of generality, the constant $k$ may be taken to be equal
to $0, 1$ or $-1$, in which case the metric $d{\bar s}^2$ describes 
flat, spherical, or hyperboloidal spatial sections respectively.
In (\ref{metr}), $m$ runs over $q$ dimensions so that $D = 1 + p + q$. 

   In the gauge $U = pA + qB$, the action (\ref{ac1}) reduces to 
\bea
S = \int dt &[&(\dot\Phi)^2 + {2q(D-2)a^2\over {p-1}}{\dot Y}^2
-{2p\Delta\over{p-1}}{\dot X}^2 \nonumber\\
&-& \Delta \lambda^2 e^{\Phi}
+ {2kp\Delta (p-1)}e^{2X}] \ .
\label{ac2}
\eea
In  writing down this action, one can us either of two ans\"atze
for the field strength $F_n$ that are compatible with
the symmetries of the metric (\ref{metr}), giving rise to electric
or magnetic cosmological solutions. In the electric solutions,
the ansatz for the antisymmetric tensor is given in terms of its
potential, and in a coordinate frame takes the form
\be
A_{m_1m_2...m_q} = f\epsilon_{m_1m_2...m_q} \ ,
\label{pot}
\ee
and hence
\be
F_{0m_1m_2...m_q} = {\dot f} \epsilon_{m_1m_2...m_q } \ ,
\label{field1}
\ee
where $f$ depends on $t$ only. For the electric solutions, we have 
$p = D-n, q = n-1$.
For the magnetic cosmological solutions, the ansatz for the 
tangent-space components for the antisymmetric tensor is
\be
F_{a_1a_2...a_p} = \lambda e^{-pA}\, \epsilon_{a_1a_2...a_p} \ ,
\label{field2}
\ee
where $\lambda$ is a constant. Thus we have $p = n, q = D-n-1$.
In the action (\ref{ac2}), $X, Y$ and $\Phi$ are related to
$A, B$ and $\phi$ in the following way:
\bea
&X& \equiv qB + (p-1)A ,\nonumber\\ 
&Y& \equiv B + {p-1\over {\epsilon 
a(D-2)}}\phi ,\nonumber\\
&\Phi& \equiv -\epsilon a\phi + 2qB \ .
\label{redef}
\eea
Here $\epsilon =1$ is for electric case and $\epsilon =-1$ for
the magnetic case. Note that in the electric case, the constant
$\lambda$ arises as the integration constant for the function $f$
in (\ref{field1}).

The equations of motion for $X, \Phi$ and $Y$ are
\bea
&&\ddot X + k(p-1)^2e^{2X} = 0 \ ,\nonumber\\
&&\ddot \Phi + {1\over 2}\Delta \lambda^2 e^{\Phi} = 0 \ ,\nonumber\\
&&\ddot Y = 0 \ .
\label{eqom}
\eea
The variation of the action (\ref{ac1}) with respect to the
lapse function ${\sqrt{g^{\phantom{\Sigma\Sigma}}_{00}}}$ provides the 
canonical constraint:
\bea
\dot \Phi^2 &+& \Delta \lambda^2 e^{\Phi} + 
{2q(D-2)a^2\over {p-1}}{\dot Y}^2 \nonumber \\
&=& {2p\Delta\over{p-1}}{\dot X}^2 +  {2kp\Delta (p-1)}e^{2X}.
\label{cons}
\eea 
Since $X$ and $\Phi$ both satisfy Liouville equations, it is 
straightforward to solve these equations directly: 
\bea
e^{-X} &=&\cases{\fft{p-1}{{\tilde P}_X}\, \cosh({\tilde P}_Xt)\ , 
& if $k=1$;\cr \fft{p-1}{{\tilde P}_X}\, \sinh({\tilde P}_Xt)\ , 
& if $k=-1$;\cr}\nn\\
X&=& -{\tilde P}_Xt\ ,\quad \hbox{if $k=0$}, \label{cases}
\eea
where ${\tilde P}_X$ is an arbitrary constant.
Similarly the solution for $\Phi$ is
\be
e^{-\ft12\Phi}= \fft{\lambda \sqrt\Delta}{2{\tilde P}_\Phi}\, 
\cosh({\tilde P}_\Phi t)
\ ,\label{psol}
\ee
where ${\tilde P}_\Phi$ is again constant. 
 The solution for $Y$ may be taken
to be simply
\be
Y= -{\tilde P}_Y t\ .\label{ysol}
\ee
The constraint (\ref{cons}) therefore implies that
\be
{{\tilde P}_\Phi}^2= \fft{p\Delta {{\tilde P}_X}^2-q(D-2) a^2 
{{\tilde P}_Y}^2}{2(p-1)}\ .\label{beta} \ee
The Hamiltonians for the fields $X$, $\Phi$ and $Y$ are given by
\bea
H_X &=& {1\over 2}P_X^2 + {1\over 2}k(p-1)^2e^{2X} \ , \nonumber\\
H_\Phi &=&{1\over 2} P_\Phi^2 + {1\over 2}\Delta 
\lambda^2 e^\Phi \ , \nonumber\\
H_Y &=& {1\over 2}P_Y^2 \ ,
\label{hamil}
\eea
where $P_X, P_\Phi$ and $P_Y$ correspond to momenta conjugate
to $X, \Phi$ and $Y$ coordinates. 
Notice that the solutions for $X, \Phi$ and $Y$ can be cast in a 
different form in terms of their phase-space variables. For example,
when $k = 1$, the solution for $X$ can be written as 
\bea
&&e^{-X} = {p-1\over {\tilde P_X}}\cosh {\tilde X},\nn \\
&&{ P}_X = -{\tilde P}_X \tanh {\tilde X} \ ,
\label{xp}
\eea
where $\tilde X = \tilde P_X t$. In fact, these equations can be 
viewed as a canonical transformation from the interacting Liouville 
system, with phase-space coordinates $(X, P_X)$, to a free system with 
phase-space coordinates ${\tilde X, \tilde P_X}$ with the Hamiltonian
${\tilde H}_X = {1\over 2}{\tilde P}_X^2$, by re-writing (\ref{xp}) as
\bea
&&P_X = (p-1)e^X \sinh {\tilde X}\nn \\
&&{\tilde P}_X = (p-1)e^X \cosh{\tilde X} \ ,
\label{xpp}
\eea
The generating function $F(X, \tilde X)$ has the following
form
\be
F(X, \tilde X) = (p-1)e^X \sinh \tilde X \ ,
\ee
such that 
\be
P_X = {\partial F\over {\partial X}} \ , \quad 
\tilde P_X = {\partial F\over {\partial {\tilde X}}}\ .
\ee
These are the same equations as in (\ref{xpp}).
Obviously, since
$H_\Phi$ has also the same structure, a similar set of canonical
transformations will also bring it to a free Hamiltonian form. 
Thus by solving a set of free systems and using the canonical 
mapping (\ref{xp}), we can generate the solutions of the interacting
theory given by the action (\ref{ac1}).
As we shall discuss in section III, these transformations can be 
implemented in the quantum version of the model. This, in turn, will 
allow us to solve the corresponding WDW equations in a straightforward
manner.
 
%%%%%%%%%%%%%%%%%%%%%%%%%%%%%%%%%%%%%%%%%%%%%%%%%%%%%%%%%%%

\subsection{Multi-charge cosmological models}

    Multi-charge solutions in $D$-dimensional maximal supergravity can be
described by the truncated action
\be
S = \int d^Dx \sqrt{-g}[ R - {1\over 2}(\partial \vec \phi )^2 -
{1\over{2n!}}\sum_{\a=1}^{N} e^{\vec c_\a \cdot \vec \phi}\, F_\a^2] \ ,
\label{ac3}
\ee
when the dilaton vectors for the set of $N$ field strengths 
$F_\a$ of rank $n\ge 2$ satisfy the dot products \cite{lpsol}
\be
M_{\a\beta} = \vec c_\a \cdot \vec c_\beta =
4 \delta_{\a\beta} - \fft{2(n-1)(D-n-1)}{D-2}\ .\label{mdot1}
\ee
The maximum value $N_{\rm max}$ for  $N$ depends on the rank of the field
strengths, and on the dimension $D$.  For example for 2-form field
strengths, $N_{\rm max}= 2$ for $6\le D \le 9$;  $N_{\rm max}=3$ in $D=5$;
and $N_{\rm max} = 4$ in $3\le D\le 4$ \cite{lpsol}.  
As before, we define fields
\bea
&X& \equiv qB + (p-1)A ,\nonumber\\
&Y& \equiv B + {p-1\over {\epsilon
a(D-2)}}\sum_{\alpha, \beta}(M^{-1})_{\alpha \beta}\varphi_\beta
\ ,\nonumber\\
&\Phi_\alpha & \equiv -\epsilon \varphi_\alpha + 2qB \ . 
\label{redef2}
\eea
where $\varphi_\a = \vec c_\a\cdot \vec \phi$.
The solutions of the equations of motion that follows from the 
 action (\ref{ac3}) in terms of these fields are (see 
\cite{lmpw} for details)
\bea
&&e^{-X} =\cases{\fft{p-1}{{\tilde P}_X}\, \cosh({\tilde P}_Xt)\ ,
& if $k=1$;\cr \fft{p-1}{{\tilde P}_X}\, \sinh({\tilde P}_Xt)\ ,
& if $k=-1$;\cr}\nn\\
&&X= -{\tilde P}_Xt\ ,\quad \hbox{if $k=0$} \ ,\nn\\
&&e^{-\ft12\Phi_\alpha} = \fft{\lambda_\alpha \sqrt\Delta}{2
{\tilde P}_{\Phi_\alpha}}\,
\cosh({\tilde P}_{\Phi_\alpha} t) \ ,\nn\\
&&Y = -\tilde P_Yt \ , 
\label{psol2}
\eea
where $\lambda_\alpha$ is the charge associated to the form field 
$F_\alpha$ and ${\tilde P}_X$, ${\tilde P}_{\Phi_\alpha}$, ${\tilde P}_Y$ 
are constants satisfying the following constraint:
\be
\sum_\alpha{{\tilde P}_{\Phi_\alpha}}^2= \fft{2p\Delta {{\tilde 
P}_X}^2-2q(D-2) a^2 {{\tilde P}_Y}^2}{\Delta (p-1)}\ .
\label{beta2} \ee
Here $\Delta = 4/N$ and $a^2$ is given in (\ref{avalue}).
The Hamiltonian that follows from (\ref{ac3}) is given by
\be
H = \sum_\alpha H_{\Phi_\alpha} + {8q(D-2)a^2\over {(p-1)\Delta}} 
H_Y - {8p\over{p-1}}H_X \ ,
\label{hamil2}
\ee
where 
\be
H_{\Phi_\alpha} = {1\over 2}P_{\Phi_\alpha}^2 + 
2\lambda_\alpha^2 e^{\Phi_\alpha} 
\ . \label{comp}
\ee
As in the previous subsection the Hamiltonian (\ref{hamil2})
can be brought to a free Hamiltonian, by means of a set of
$N+1$ canonical transformations which act on the phase
space variables $(\Phi_\alpha, P_{\Phi_\alpha})$ and $(X, P_X)$.

%%%%%%%%%%%%%%%%%%%%%%%%%%%%%%%%%%%%%%%%%%%%%%%%%%%%%%%%%%%

\subsection{$SL(3, R)$ Toda cosmological models}

As discussed in \cite{lmpw}, when the space-time dimension is $D = 2n$,
the $n$-rank field strength can carry both electric (\ref{field1})
and magnetic (\ref{field2}) charges. In this case, $p =n$
and $q = n-1$. Making the same gauge choice as before, the equations
of motion that follow from the action (\ref{ac1}) reduce to
\bea
&&\ddot X + k(n-1)^2 e^{2X} =0\ ,\nonumber\\
&&\ddot q_1 = - e^{\a q_1 + (1-\a) q_2}\ ,\nonumber\\
&&\ddot q_2 = - e^{(1-\a) q_1 + \a q_2}\ ,
\label{toda1}
\eea
where 
\bea
X&=& (n-1) (A + B)\ ,\nonumber\\
B&=& \fft1{4(n-1)} \Big( q_2 + q_1 -2 \log((n-1)\lambda_1\lambda_2)\Big)\ ,\\
\phi &=& \fft{a}{2(n-1)} (q_2 - q_1) + \fft1{a}
\log\fft{\lambda_1}{\lambda_2}\ , \nonumber
\eea
and the constant $\alpha$ is given by
\be
\a = \ft12 + \fft{a^2}{2(n-1)} = \fft{\Delta}{2(n-1)}\ .
\ee
The first-order constraint in this case reduces to
\bea
\ft12\a(\dot q_1^2 + \dot q_2^2) &+& (1-\a) \dot q_1 \dot q_2
+ e^{\a q_1 + (1-\a) q_2} + e^{\a q_2 + (1-\a) q_1} \nonumber\\
&=&
2n\Big( \dot X^2 +  k(n-1)^2 e^{2X} \Big)\ .
\eea
In (\ref{toda1}), $\lambda_1$ and $\lambda_2$ correspond to 
electric and magnetic charges. In particular, the choice 
$\lambda_1 = \lambda_2$ will correspond to a self-dual 
cosmological model.  Although for a generic value of $\alpha$
the equations seem not to be integrable, when $\alpha = 2$ equation 
(\ref{toda1}) reduces to the $SL(3, R)$
Toda equations, which can be exactly solved. This value of $\alpha$ can
arise for a 2-form field strength in $D = 4$, with $\Delta =4$.
The Hamiltonian of the $(q_1,q_2)$ system can be writtrn as
\be
H_{(q_1,q_2)} = {1\over 3}(P_1^2 + P_2^2 + P_1P_2)
+ e^{2q_1-q_2} + e^{2q_2 -q_1} \ ,
\label{todah}
\ee
where $P_1, P_2$ are the momenta given by
\be
P_1 = 2\dot q_1 - \dot q_2 \ , \quad P_2 = 2\dot q_2 - q_1 \ .
\ee
As in the Liouville case, there exists a set of canonical 
transformations which maps the above Toda Hamiltonian to
a free Hamiltonian. The mapping is given in \cite{anps}:
\bea
e^{-q_1} &=& 
{e^{\tilde q_1}\over {\tilde P_1 (\tilde P_1 -
\tilde P_2)}} +
{e^{\tilde q_2}\over {\tilde P_2 (\tilde P_1 -
\tilde P_2)}} +
{e^{(-\tilde q_1 - \tilde q_2)}\over {\tilde P_1\tilde P_2}} \nn \\
e^{-q_2} &=& 
{e^{-\tilde q_1}\over {\tilde P_1 (\tilde P_1 -
\tilde P_2)}} +
{e^{-\tilde q_2}\over {\tilde P_2 (\tilde P_1 -
\tilde P_2)}} +
{e^{(-\tilde q_1 - \tilde q_2)}\over {\tilde P_1\tilde P_2}} 
\label{qq}
\eea
and
\bea
(2P_1 + P_2) e^{-q_1} &=& -
{(2\tilde P_1 -\tilde P_2) e^{\tilde q_1}\over{\tilde P_1(\tilde P_1 -
\tilde P_2)}} -
{(2 \tilde P_2 - \tilde P_1)e^{\tilde q_2}\over 
{\tilde P_2 (\tilde P_1 - \tilde P_2)}} \nn\\
&\quad +&
{(\tilde P_1 + \tilde P_2) e^{(-\tilde q_1 - \tilde q_2)}\over {\tilde 
P_1\tilde P_2}} \nn \\
(2P_2 + P_1) e^{-q_2} &=&
{(2\tilde P_1 -\tilde P_2) e^{-\tilde q_1}\over{\tilde P_1(\tilde P_1 -
\tilde P_2)}} +
{(2 \tilde P_2 - \tilde P_1)e^{-\tilde q_2}\over 
{\tilde P_2 (\tilde P_1 - \tilde P_2)}} \nn\\
&\quad -&
{(\tilde P_1 + \tilde P_2) e^{(\tilde q_1 +\tilde q_2)}\over {\tilde 
P_1\tilde P_2}}.
\label{pp}
\eea
With this transformation, in terms of the new variables, the
Toda Hamiltonian reduces to a free Hamiltonian of the form
\be
H_{(\tilde q_1, \tilde q_2)}  = {1\over 3} (\tilde P_1^2 + 
\tilde P_2^2 - \tilde P_1 \tilde P_2 )\ .
\ee

%%%%%%%%%%%%%%%%%%%%%%%%%%%%%%%%%%%%%%%%%%%%%%%%%%%%%%%%%%%%

\subsection{$E_{N}$ Toda cosmological models}

             Maximal supergravities in $D$ dimensions coming from the
toroidal compactification of eleven-dimensional supergravity have
$E_{11-D}$ global symmetries.  It is natural therefore to expect that
there might exist $p$-brane or cosmological solutions that arise as
solutions of the $E_{N}$ Toda equations.  It has been observed that
the dilaton vectors for all the axions are precisely in one-to-one
correspondence with the positive roots of the $E_{11-D}$ algebra.  In
particular, the simple roots can be taken to be $\vec b_{i,i+1}$ and
$\vec a_{123}$ \cite{cjlp}.  Thus in all dimensions we may summarise the
information about the dot products of the dilaton vectors for the full
sets of axions by the Dynkin diagram:

\bigskip\bigskip

\centerline{
\begin{tabular}{ccccccccccccc}\\
 $\vec b_{12}$& &$\vec b_{23}$& &$\vec b_{34}$& &$\vec b_{45}$
& &$\vec b_{56}$& &$\vec b_{67}$& &$\vec b_{78}$ \\
 o&---&o&---&o&---&o&---&o&---&o&---&o\\
 &   & &   &$|$&   & &   & &   & &   & \\
 &   & &   &o&   & &   & &   & &   & \\
 &   & &   &$\vec a_{123}$& & &   & &   & &   & \\
\end{tabular}}
\bigskip\bigskip

\centerline{Table 1: The dilaton vectors $\vec b_{i,i+1}$ and $\vec
a_{123}$}
\centerline{generate the $E_n$ Dynkin diagram}
\bigskip\bigskip
In each dimension $D$, the diagram is truncated to the part that survives
when only the simple roots with indices $i\le 11-D$ are retained.

          It is straightforward to verify that when the axions take
the standard electric or magnetic ans\"atze, the full Lagrangian
can be consistently truncated to one of the form (\ref{ac3}) with the
$N$ field strengths $F_\a=(F_0^{(123)}, {\cal F}_0^{(12)}, 
{\cal F}_0^{(23)},
\ldots)$, and associated dilaton vectors given by $\vec c_\a =
(\vec a_{123}, \vec b_{12}, \vec b_{23}, \ldots)$.   Now the dilaton
dot produts $M_{\a\beta}$ are no longer given by (\ref{mdot1}); instead 
$M_{\a\beta}$ is precisely the Cartan matrix for $E_{N}$. We shall now
show
that this has the consequence that the equations of motion of the
system can be cast into the form of the one-dimensional $E_{N}$ Toda
equations.   To do this, we first consistently truncate the Lagrangian
further to
\bea
e^{-1} {\cal L} &=& R -\ft12 \sum_{\a,\beta=1}^N (M^{-1})_{\a\beta}
\del_{\sst M} \varphi_\a \del^{\sst M} \varphi_\beta \nn\\
&&-\ft12 \sum_{\a=1}^N e^{\varphi_\a} (\del \chi_\a)^2\ ,\label{enlag}
\eea
where $\varphi_\a = \vec c_\a \cdot \vec \phi$.
We shall discuss the electric solutions in detail, for which $p=D-1$
and $q=0$.  (The discussion for the magnetic solutions is analogous.) 
The metric ansatz in this case is thus given by
\be
ds^2 = -e^{2U} dt^2 + e^{2A} d\bar s^2\ ,
\ee
where $d\bar s^2$ is again the metric on the spatial sections,
typically taking the form (\ref{sec}).  It is convenient to make the
gauge choice $U=(D-1)A$, which implies that the equations of motion can be
written as
\bea
&&\ddot A + k (D-2) e^{2(D-2)A} =0\ ,\label{aeom}\\
&&\ddot \Phi_\a = -\lambda_\a^2 \exp(\ft12 \sum_\beta M_{\a\beta}
\Phi_\beta)\ ,\label{entoda1}
\eea
where $\Phi_\a = -2\sum_\beta (M^{-1})_{\a\beta} \,\varphi_\beta$,
together with a first-order equation
\bea
&&\ft14\sum_{\a\beta}M_{\a\beta}\, \dot \Phi_\a \dot \Phi_\beta +
\sum_a\lambda_\a^2 \exp(\ft12 \sum_\beta M_{\a\beta} \Phi_\beta)
\nn\\
&&=2 (D-2)(D-1) (\dot A^2 + ke^{2(D-2)A})\ .\label{foen}
\eea
Defining $\Phi_\a = q_\a -4 \sum_\beta (M^{-1})_{\a\beta}
\log(\lambda_\beta)$ to remove the charges, the
equations for $\Phi_\a$ then become
\bea
&&\ddot q_1 = -e^{2q_1 - q_4}\ ,\qquad\qquad
\ddot q_2 = -e^{2q_2 - q_3}\ ,\nn\\
&& \ddot q_3 = -e^{-q_2 + 2 q_3 - q_4}\ ,\qquad
\ddot q_4 = -e^{-q_1 - q_3 + 2 q_4 - q_5}\ ,\nn\\
&&\ddot q_5 = -e^{-q_4 + 2 q_5 - q_6}\ ,\qquad
\ddot q_6 = -e^{-q_5 + 2 q_6 - q_7}\ ,\label{e8}\\
&&\ddot q_7 = -e^{-q_6 + 2 q_7 - q_8}\ ,\qquad
\ddot q_8 = -e^{-q_7 + 2 q_8}\ .\nn
\eea
These are precisely the $E_8$ Toda equations.  Here we present only the
$E_8$ case, since the lower cases for $E_N$ with $N\le 7$ are obtained
by straightforward truncation.  (In fact an alternative truncation can 
instead be made that reduces the $E_8$ Toda equations to the $SL(N+1,\R)$
equations that were discussed previously \cite{lpsln}.)
The left-hand side of the first-order equation
(\ref{foen}) is the Hamiltonian for the Toda equations (\ref{e8}),
given by
\be
{\cal H} = \ft14 \sum_{\a\beta} M_{\a\beta}\, \dot q_a \dot q_\beta
+ \sum_\a \exp(\ft12 \sum_\beta M_{\a\beta}\, q_\beta)\ .\label{cham}
\ee
The right-hand-side of the equation (\ref{foen}) is a constant,
given by $2(D-1) P_X^2/(D-2)$, since the function A, satifying
(\ref{aeom}), is given by (\ref{cases}) with $X=2(D-2)A$.  
Thus the first-order equation
(\ref{foen}) means no more than that the Hamiltonian is a conserved
quantity, given by
\be
{\cal H} = \fft{2(D-1)P_X^2} {D-2}\ .
\ee
%%%%%%%%%%%%%%%%%%%%%%%%%%%%%%%%%%%%%%%%%%%%%%%%%%
\section{The Wheeler-DeWitt Equation}

Recently, there have been attempts to solve the WDW
equation in string cosmology and study its implications. Let us now
construct the WDW equations for the string cosmological models that we
are considering in this paper.  We recall that the classical equations
of motion, which correspond to interacting Liouville or Toda systems,
can be reduced to free field equations after implementing the
canonical transformations discussed in the previous section. In fact
it has been shown that these classical transformations can be extended
to the quantum level. This is achieved by introducing intertwining
operators, which implement the canonical transformations on the
quantum mechanical operators and wave functions \cite{anps}. In what
follows, we explicitly construct the intertwining operators for the
cosmological models that have been discussed in sections IIA and IIB,
and we use these to obtain solutions of the corresponding WDW
equations. We end this section with the analysis of some of the
solutions of the WDW equations, by imposing proper boundary conditions
on the wave functions.

%%%%%%%%%%%
\subsection{Intertwining operators and the solutions of WDW equation}

The canonical transformation between the classical Liouville and free
theories that have been discussed in the previous section can be
implemented at the quantum level. This is done by introducing
intertwining operators \cite{anps} which generate canonical
transformations on the quantum operators and on the wave functions. In
order to construct such operators we first focus on the simplest of
all the models that have analysed in section II, namely, the
cosmological model with a single charge.

In order to proceed, let us first concentrate on $H_X$ given in
(\ref{hamil}).  It is known that there exists an operator
$C_X$ which transforms the Liouville Hamiltonian to a free one \cite{anps}.  
In particular,
\be
C_X H_X C_X^{-1} = {\tilde H_X} \ .
\label{freeh}
\ee
As a result, the wave functions $\psi_X$ and $\tilde \psi_X$
of the Liouville and free theories are related by
$\psi = C_X^{-1} \tilde \psi$.
The operator $C_X$ has been constructed in \cite{anps}, and takes the 
following form:
\be
C_X = {\cal P}_{(p-1)\sinh X} P_X^{-1} {\cal I} {\cal P}_{\ln X} \ ,
\label{opc}
\ee
where each of the constituent pieces has the following action:
\bea
&&{\cal P}_{\ln X} : \quad X \rightarrow \ln X \ , \quad P_X \rightarrow
XP_X \ ,\nn\\
&&{\cal I}_X :  \quad X \rightarrow P_X  \ ,  \quad P_X \rightarrow - X \
\ ,\nn\\ && P_X^{-1} : \quad  X \rightarrow P_X^{-1}X P_X \ , \quad P_X
\rightarrow P_X \ ,\nn\\
&&{\cal P}_{(p-1)\sinh X} : \quad X \rightarrow (p-1)\sinh X \ ,
\nn\\
&& \qquad \qquad \qquad \quad P_X \rightarrow {P_X {\cosh}^{-1} X\over
{p-1}} \ . \label{trans}
\eea
Taking into account the commutation relation $[P_X, X] = -i$, it
is immediate that the combined action of (\ref{trans}) is
to map the Liouville Hamiltonian $H_X$ to the free Hamiltonian 
$\tilde H_X=\ft12 \tilde P_X^2$. Similarly,
the operator $C_X$ has the following action  on the wave function 
\cite{anps}:   
\be
C_X^{-1} : \quad e^{ikX} \rightarrow N_k\, K_{ik}(e^X) \ ,
\label{wavefun}
\ee
where $K_{ik}$ is a modified Bessel function.  Owing to the fact that the
canonical transformation described by $C_X$ is non-unitary (as it must be,
since the Liouville theory is not simply equivalent to the free theory),
the normalisation of the transformed wave function is not just the same as
the normalisation of the free wave function.  It can be determined by 
calculating the effect of the transformation on the Hilbert-space inner
product, leading to the result \cite{anps}
\be
N_k = \fft1{\pi}\, \sqrt{2k\, \sinh(\pi k)}\ .\label{norm}
\ee

Now consider the WDW equation, which is simply
\be
H \Psi (X, \Phi, Y) = 0 \ .
\label{wdwone}
\ee
Here the total Hamiltonian of the system is given by
\be
H = H_\Phi + {2q(D-2)a^2\over {p-1}} H_Y -{2p\Delta\over {p-1}}H_X \ .
\label{totalh}
\ee
It is clear now from the structure of the Hamiltonian that the
wave function $\Psi (X, \Phi, Y)$ will have the following form:
\be
\Psi (X, \Phi, Y) = \Psi_X \Psi_\Phi e^{iP_Y Y} \ ,
\label{si}
\ee
where $\Psi_X$ and $\Psi_\Phi$ depend on $X$ and $\Phi$ respectively.
Following our previous discussion, 
there is an intertwining operator which will convert the interacting
Hamiltonian $H$ to a sum of free Hamiltonians. It is given by
\be
C = {\cal P}_{(p-1)\sinh X} P_X^{-1} {\cal I}_X {\cal P}_{\ln X}
 {\cal P}_{{\sqrt\Delta}\lambda\sinh \Phi}
P_\Phi^{-1} {\cal I}_\Phi {\cal
P}_{\ln \Phi} \ . \label{inter1}
\ee
Its action on the Hamiltonian is
\be
C H C^{-1} = {\tilde H}_\Phi +
{2q(D-2)a^2\over {p-1}} H_Y -{2p\Delta\over
{p-1}}{\tilde H}_X \ .
\label{free}
\ee
It is now easy to read off the action of $C$ on the wave functions:
\bea
\Psi (X, Y, \Phi) &=& \fft1{\sqrt{2\pi}} N_{k_X} N_{k_\Phi} \nn \\
&\times&K_{i(p-1)k_X}(e^X)
K_{i{\sqrt\Delta}\lambda k_\Phi}(e^\Phi)e^{ik_Y Y} \ ,
\label{sif}
\eea
where $N_{k_X}$ and $N_{k_\Phi}$ are momentum-dependent
normalisation constants which can be determined from (\ref{norm}).

So far, we have been discussing the case $k=1$. Following similar
arguments, we can also study the WDW wave function for an open
universe, for which $k = -1$. In this case, the analogue of   
(\ref{trans}) is
\bea
&&{\cal P}_{\ln X} : \quad X \rightarrow \ln X \ , \quad P_X \rightarrow
XP_X \ ,\nn\\
&&{\cal I}_X :  \quad X \rightarrow P_X  \ ,  \quad P_X \rightarrow - X \
,\nn\\ && P_X^{-1} : \quad  X \rightarrow P_X^{-1}X P_X \ , \quad P_X
\rightarrow P_X \ ,\nn\\
&&{\cal P}_{(p-1)\cosh X} : \quad X \rightarrow (p-1)\cosh X \ ,  
\nn\\
&& \qquad \qquad \qquad \quad P_X \rightarrow {P_X {\sinh}^{-1} X\over 
{p-1}} \ . \label{trans2}  
\eea
The operator $C$ is now
\be
C_X = {\cal P}_{(p-1)\cosh X} P_X^{-1} {\cal I} {\cal P}_{\ln X} \ ,
\label{opc1}
\ee
whose action on the wave functions can be evaluated using similar methods
to those decsribed in \cite{anps}, which we used above in the $k=1$ case.

We shall not discuss the $k=0$ case in detail. Following the above
discussion, the structure of the wave function is also easily
obtained in this case.
%%%%%%%

Consider now the multi-charge cosmological models discussed in section
IIB.  As we saw, by proper choice of variables the Hamiltonian
can be brought to the form of a sum of $N+1$ Liouville equations,
together with a free part, as given in equation (\ref{comp}).  Thus,
following the above discussion, one can immediately construct the
quantum intertwining operator $C$ for this case. For $k = 1$, it is
\bea
C &=& {\cal P}_{(p-1)\sinh X} P_X^{-1} {\cal I}_X {\cal P}_{\ln X}  \nn\\
&\times &
\prod_\alpha {\cal P}_{{{\sqrt\Delta} \lambda}\sinh {\Phi_\alpha}}
P_{\Phi_\alpha}^{-1}
{\cal I}_{\Phi_\alpha} {\cal P}_{\ln {\Phi_\alpha}} \ .
\label{inter2}
\eea
The action of $C$ on the wavefunction $\Psi$ is again easily read
off:
\bea
\Psi (X, Y, \Phi) = && \fft1{\sqrt{2\pi}} N_{k_X} K_{i(p-1)k_X}(e^X)
e^{ik_Y Y}\nn\\
&&\times\prod_\alpha
N_{k_{\Phi_\alpha}}K_{i{\sqrt\Delta}\lambda
k_{\Phi_\alpha}}(e^{\Phi_\alpha}) \ , \label{sif2}
\eea
and the normalisation constants $N_{k_X}$ and $N_{k_{\Phi_\a}}$ 
can be determined from (\ref{norm}).

We shall not discuss the $k = -1$ and $k = 0$ cases seperately here,
since the wave functions can be obtained easily by following the
previous discussion. We should like to mention here that 
for the case of the $SL(3, R)$ Toda model, the intertwining operator 
can also be constructed, by generalising the transformation
of (\ref{qq}) at the operator level \cite{anps}. The corresponding
wave function can also be computed.  
%%%%%%%%%

\subsection{Analysis of WDW wave functions}

Here we analyse some of the solutions of the WDW equation discussed in
the previous subsection, by imposing proper boundary conditions on the
wave functions.

In order to study the solutions of the WDW equation obtained above,
we first note that it is necassary to specify the intial boundary
conditions.  When we look at the classical cosmological solutions
given in Sec II, we see from (\ref{cases}) - (\ref{ysol}), that one
has to specify $p$, ${\tilde P}_X$, $\Delta$, $a$, ${\tilde P}_{\Phi}$
and ${\tilde P}_Y$.  Furthermore, we have to choose the value $k=-1$,
0, or 1.  We shall present two specific cases to illustrate how we can
obtain explicit solutions, and then discuss their properties.

Let us consider the string effective action in $D=10$, such that
$p=3$, $q=6$, $a^2=1$ and $\Delta=4$. Furthermore, we look at the
magnetic sector of this NS-NS case, and so $\epsilon=-1$; see
equations (\ref{ac1}) - (\ref{redef}) for definitions of the
parameters we specified above. We shall take $k=0$ from now on.  The
solutions correspond to
\be
X=-{\tilde P}_Xt,~Y=-{\tilde P}_Yt,~
e^{-2\phi}= {\lambda \over{{{\tilde P}_{\Phi}}}} 
\cosh({\tilde P}_{\Phi}t +\gamma)\ ,
\ee
where $\phi$ is the dilaton. Note that for the case at hand, the coupling
constant of the theory is identified to be $g_s = e^{-\phi}$. We 
recall that the integration constants
satisfy the constraint
\bea {{\tilde P}_{\Phi}}^2 = 3{\tilde P}_X^2 - 12 {\tilde P}_Y^2 \eea
as is evident from eq.(15). Instead of examining the two parameter problem,
(note that we can set $\gamma=0$ without any loss of generality and keep $
\lambda$ as an arbitrary parameter) let us look at two  cases separately with
specific choices for the value of the parameters.
\bigskip

\noindent CASE I:

\bigskip

Let us first consider the case when ${\tilde P}_Y = 0$ and ${\tilde P}_X <0$.
Then it follows that ${\tilde P}_X =-{1\over{\sqrt 3}} {\tilde P}_{\Phi} 
$ and $U=3A+6B$, where
\bea 
e^{{8\over 3}A}&=& {\lambda \over{{\tilde P}_{\Phi}}}{\rm cosh}
({\tilde 
P}_{\Phi}t ) e^{{4\over 3}X} \nn \\
e^{-8B} &=& {\lambda\over{{\tilde P}_{\Phi}}}{\rm cosh}({\tilde P}_{\Phi}t   
) e^{{4\over 3}X}
\eea
Now, let us examine the behaviour of $e^U$, $e^A$ and $e^B$ 
for $t\rightarrow \pm \infty$
\bea e^U \rightarrow e^{{{3\over 8} {\tilde P}_{\Phi} {\vert t \vert} + {1\over
{\sqrt 3}} t}} 
\eea
We see that as $t \rightarrow +\infty $, $e^U \rightarrow +\infty$ and
as $t \rightarrow -\infty$, $e^U \rightarrow 0$. We can define a
comoving time as follows: $d \tau = e^U dt$. Therefore, $0 \le \tau
\le \infty$ since $-\infty \le t \le \infty$. We can define two scale
factors, $R_A$ and $R_B$ respectively, as $R_A = e^A$ and $R_B=
e^B$. Notice that for large $\tau$ $R \rightarrow {\tau}^{\alpha}, 0<
\alpha <0$. We also note that for $\tau \rightarrow 0$ and $\tau
\rightarrow \infty$ this scale factor tends to $\infty $. The other
scale factor $R_B$ tends to zero in these two limits.  Note
that $R'_A\ge 0$ and $R'_B \le 0$ in this case. Since
$e^{-\phi}$ is the coupling constant for this magnetically-charged
case, we see that for $\tau =0$ and $\tau =\infty$, we end up in the
strong coupling phase.

Let us look at the wave function obtained from the solution of the WDW 
equation.  The Bessel functions of our choice are:
\be 
\Psi(X,Y, \Phi) = N_X N_YN_{\Phi} e^{iP_X X} K_{2i\lambda 
P_{\Phi}}(e^{\Phi})\ ,  
\ee
where the normalisation constants are determined from (\ref{norm}) as
usual. We recall that for $t \rightarrow \pm \infty$, 
$\Phi \rightarrow 0$; and from 
the relation between $\Phi$ and dilaton $\phi$, we also know that the 
coupling constant $e^{-\phi}$ diverges in this limit. The wave function
\bea
\Psi(X,Y, \Phi) =&& N_XN_YN_{\Phi} \Gamma (2i\lambda P_{\Phi}) \nn \\
&&e^{2i\lambda P_{\Phi} ln 2}
 e^{P_XX}e^{-2i\lambda P_{\Phi} \Phi} \eea
is obtained in the limit when the scale factor $A$ tends to large values and
$\Phi$ tends to zero.

\bigskip

\noindent CASE II:
\bigskip

Let us consider another interesting case when $P_Y$ is non-zero
and $P_{\Phi} = 6P_Y$. The constraint equation implies that $P_X =
\pm 4P_Y$, for which we shall choose the plus sign.  Note that for this 
choice we have
\be 
e^A \rightarrow e^{5 P_Y t},\quad  e^B \rightarrow e^{-P_Yt}, \quad
e^U \rightarrow e^{{21 \over 2} t}\ , 
\ee
as $t \rightarrow \infty$. In the limit of $t \rightarrow - \infty$, we have
\be 
e^A \rightarrow e^{{1\over 2}P_Yt}, \quad
e^B \rightarrow e^{{1\over 2}P_Y t}, 
\ee
and it is easy to see that $R'_A$ is positive and $R''_A$ is negative. 
When we
consider the wavefunction for this case, namely
$ \Psi (X, Y, \Psi)$, in the limit when $\Phi \rightarrow 0$, it has the form
\bea 
\Psi(X, Y, \Phi) = &&N_XN_YN_{\Phi} e^{4iP_Y X}e^{iP_Y}
e^{12i\lambda P_Y ln2} \nn \\
&&\Gamma (12i \lambda P_Y)e^{-12i\lambda P_Y \Phi} \ .
\eea

\bigskip

\noindent CASE III:

\bigskip
Now we consider the situation when there is non-zero spatial curvature, 
corresponding to $k = -1$. For the sake of simplicity, let us choose
$P_Y =0$ and therefore, from the constraint equation, we have 
$P_X = \pm {{1 \over {\sqrt 3}}} {P_{\Phi}}$.
 We shall again choose the plus sign. It follows from the solutions
of the equations of motion that $e^X$ has a singularily at $t=0$. If we 
extract
the two scale parameters: $R_A = e^A$ and $R_B=e^B$ we find that $R_A$ is
singular at $t=0$ and $R_B$ is not. Furthermore, $\Phi \rightarrow$ 
constant as $t \rightarrow 0$. 
Now let us look at the $t \rightarrow
\infty$ limit. We find in this case that $R_A \rightarrow \infty$,  
$R_B
\rightarrow 0$ and $\Phi \rightarrow 0$. We can extract the wavefunction in
these two limits from the behavior of the Bessel functions.
 
When $t \rightarrow 0$, the wavefunction takes the form,
\bea 
\Psi(X,Y, \Phi) = N_X\, N_YN_{\Phi}\,  {\sqrt {\pi \over {2Z}}}e^{-Z}
K_{i2\lambda P_{\Phi}}(e^{\Phi}) 
\eea
where $Z = e^X$   
Note that in this limit the argument of the Bessel function takes a finite 
value.
In the other limit, {\it i.e.}\  $t \rightarrow 0$, we find that 
$e^X$ and $\Phi$ tend to zero, and so we must take limits in both of the 
Bessel functions. Thus the wave function is
\bea 
\Psi (X, Y, \Phi) = &&N_XN_YN_{\Phi} e^{2i{1\over {\sqrt 3}} P_{\Phi}
ln 2} \nn \\
&&\Gamma (i {{2 \over {\sqrt 3}}}P_{\Phi})e^{2i\lambda P_{\Phi} ln 2}\nn \\
&&\Gamma (2i \lambda P_{\Phi})e^{2i\lambda P_{\Phi} \Phi} 
\eea

%%%%%%%%%%%%%%%%%%%%%%%%%%%%%%%%%%%%%%%%%%%%%%%%%%
\section{Generalised Brans-Dicke model}
In the cosmological context, one of the extensively discussed 
models is the Brans-Dicke (BD) model \cite{barr}. Although at late times it 
reduces to Einstein's gravity, at early times its behaviour is very 
different. This is essentially because of the presence of a
scalar which couples to the metric in a non-trivial way.
Here, we are interested in a similar model, but extended to
include different form fields. A sub-class of such models has
previously been analysed, for example in \cite{sm,l,cliw}. 
The action for the theory is given by 
\be
S = \int d^Dx {\sqrt {\tilde g}} e^{-\tilde\phi}
[\tilde R - \omega (\partial\tilde\phi )^2  + {1\over {2n!}}
e^{c\tilde\phi} {\tilde F}_n^2 ] \ ,
\label{bd}
\ee
where the constant $\omega$ is known as the BD parameter.

We would like to show here that starting from the action (\ref{ac1}),
and then properly rescaling the metric and redefining various fields,
we end up with an action of the form (\ref{bd}).  Thus the solutions
of the classical equations of motion of the generalised BD theories
can simply be obtained from our analysis in section II. Furthermore,
the solutions of the WDW equation in the mini-superspace will follow
simply from those in section III, after making the necessary field
redefinitions.

We begin with the action (\ref{ac1}), which is written in the
Einstein frame. After rescaling the dilaton according to
\be
\phi = {\sqrt{2({D-1\over{D-2}} +\omega )}}\tilde\phi \ ,
\label{scal1}
\ee
we get 
\be
S = \int d^Dx{\sqrt{g}}[ R - ({D-1\over {D-2}} +\omega)
(\partial\tilde\phi )^2
+ {1\over {2n!}}e^{b\tilde\phi} F_n^2] \ .
\ee
Here $\omega$ is a constant and $b = a{\sqrt{({D-1\over{D-2}} +
\omega )}}$. Now, after the further metric rescaling 
\be
\tilde g_{\mu\nu} = e^{{2\tilde\phi\over {D-2}}}g_{\mu\nu} \ ,
\label{scal2}
\ee
we get (\ref{bd}),
%
%\be
%S = \int d^Dx {\sqrt {\tilde g}} e^{-\tilde\phi}
%[\tilde R - \omega (\partial\tilde\phi )^2  + {1\over {2n!}}
%e^{c\tilde\phi} {\tilde F}_n^2 ] \ ,
%\label{bd}
%\ee
with $c = b + {2(n-1)\over {D-2}}$. 

As mentioned earlier, the action (\ref{bd}) has been analysed for $D
=4$ and $n =3$ in \cite{sm,l,cliw}, and its cosmological properties
have been studied. However, following our previous discussion, it is
immediate that cosmological solutions of (\ref{bd}) can be obtained by
simply taking the solutions of section II and scaling the fields as in
(\ref{scal1}) and (\ref{scal2}).  As the procedure is straightforward,
we will not carry it out here. However, we should like to make
the following comments. In most of the solutions, the dilaton field
$\phi$ becomes singular at some value of the proper time. Moreover,
since the field scalings in (\ref{scal2}) involve the dilaton, the
cosmological properties of the BD metric will be considerably
different from those seen in section II.

%%%%%%%%%%%%%%%%%%%%%%%%%%%%%%%%%%%%%%%%%%%%%%%%%%%%%%%

\section{Conclusions}

    In this paper, we have analysed some of the cosmological models
arising as solutions of the low-energy effective actions of string
theories or M-theory.  These solutions involve time-dependent metric
tensor, dilaton, and antisymmetric tensor fields. The ansatz and
conventions of Ref. \cite{lmp} were used in constructing general
solutions.  As is well known, the maximal supergravities in $D$
dimensions, which arise from the toroidal compactification of
eleven-dimensional supergravity, have $E_{11-D}$ global symmetries.
We showed that in certain cases the equations of motion can be cast in
the form of the one-dimensional $E_N$ Toda equations. This was
demonstrated explicitly for the maximal $E_8$ case.

   We then studied some of the simpler cosmological models (which are
related to the Liouville equation) at the quantum level, by obtaining
the solutions of the Wheeler-DeWitt equation in a minisuperspace
approximation.  The WDW solutions were obtained using the techniques
described in \cite{anps}. This involves constructing a
quantum-mechanical canonical transformation that maps the Liouville
theory to a free theory.  The Liouville wave functions are then
obtained from the free wave functions by means of intertwining
operators.  We presented solutions of the WDW equations in several
cases, where we discussed the initial boundary conditions and studied
the evolution of the wave functions.  We showed how the wave functions
relate different domains of the theory, and we studied their
asymptotic behaviour at large times.  We also briefly discussed
generalised Brans-Dicke theories including higher-degree field
strengths, and showed how they can be related to the our framework.
Finally, in an appendix, we studied the $E_N$ Toda equations for
$(D-3)$-brane solitions, which are closely related to the $E_N$
cosmological models, and showed how they may be solved in the extremal
limit.  The case of the $D_4=O(4,4)$ Toda equations was presented
explicitly.

\section*{Acknowledgement}

The work of C.N.P is supported in part by DOE grant DE-FG03-95ER40917
and S.M is supported by NSF grant PHY-9411543. J.M. would like to
thank Professors C. Bachas and A. Chakrabarti for their gracious
hospitality and he would like to acknowledge Jawaharlal Nehru Memorial
Fund for award of the Jawaharlal Nehru Fellowship.

\appendix

\section{Extremal $E_N$ Toda $p$-branes}

       In section 2, we showed that we can obtain cosmological
solutions supported by 1-form field strengths, whose equations of
motion can be cast into $E_N$ Toda equations.  One-form field
strengths can also support electric instanton or magnetic
$(D-3)$-branes.  In this appendix, we shall discuss extremal $E_N$
Toda instantons or $(D-3)$-branes using the same set of field strengths
discussed in section 2.  (Note that the non-extremal $p$-branes are
equivalent to cosmological solutions that we discussed in section 2, 
after certain Wick rotations are performed \cite{lmp}, and we shall not 
consider these here.)   The Lagrangian is given by (\ref{enlag}).  We shall
first consider magnetic $(D-3)$ branes, with the standard metric and
field strength an\"atze
\bea
ds^2 &=& \eta_{\mu\nu} dx^\mu dx^\nu + 2^{B(r)} (dr^2 + r^2 d\theta^2)
\ ,\nn\\
\chi_\a &=& 4Q_\a\theta\ .
\eea
The extremal condition implies that $B=-1/2 \sum_\a \Phi_\a$, with
$\Phi_\a=-2 \sum_\beta (M^{-1})_{\a\beta} \varphi_\beta$.  The
equations of motion can be then written as
\be
q_\a'' = \exp(\ft12 \sum_{\a,\beta} M_{\a\beta}\, q_\beta)\ ,
\label{toda8}
\ee
with extremality implying that its Hamiltonian
\be
{\cal H} = 4 \sum_{\a,\beta}(M^{-1})_{\a\beta}\, p_\a p_\beta
-\sum_\a \exp (\ft12 \sum_\beta M_{\a\beta}\, q_\beta)\label{pham} 
\ee
vanishes.  Here a prime denotes a derivative with respect to $\rho =
\log r$.  

    Note that the positive sign on the right-hand side of (\ref{toda8})
is the opposite of that in the cosmological equations that we discussed
previously.  Correspondingly, there is a minus sign in the second term
in the Hamiltonian (\ref{pham}), whereas in the cosmological case, the
Hamiltonian (\ref{cham}) is positive definite.  This means that unlike
in the cosmological case, here we can find simple but non-trivial solutions 
for which the Hamiltonian vanishes \cite{lpsln}.  These are in fact 
extremal $p$-branes, for which the
functions $e^{-q_\a}$ can be expanded in terms of polynomials in
$\rho$:
\be
e^{-q_\a} = \sum_{m=0}^{n_\a} a_{\a m}\, \rho^m\ .\label{poly}
\ee
The integrability of the Toda equations implies that the above
series have only finite degrees $n_\a$, which we shall now determine.  
To do this, we first consider the following simple ansatz for a 
particular, special solution:
\be
e^{-q_\a} = c_\a\, H^{n_\a}\ ,\label{abcd}
\ee
where the $c_\a$ are constants and $H=1+c\, \rho$ is a single ``harmonic''
function.  Thus we have $q_\a'' = n_\a\, c^2\, H^{-2}$, and so by 
substituting into the Toda equations (\ref{toda8}), we find that they are
all satisfied provided that the exponents $n_\a$ in (\ref{abcd}) are chosen
to be
\be
n_\a = 4 \sum_\beta (M^{-1})_{\a\beta}\ ,\label{expon}
\ee
and that the constants $c_\a$ are chosen appropriately.  Thus in this
special case we see that the highest powers $n_\a$ of $\rho$ in the 
polynomials (\ref{abcd}) are given by (\ref{expon}).  In fact this special
solution corresponds to choosing the charges $Q_\a$ to occur in a certain
fixed ratio for which the solution reduces to a single-scalar one 
\cite{lpsol}.  More generally, if we relax this latter condition, we get
solutions of the form (\ref{poly}) for which the number of free parameters
is equal to the number of charges \cite{lpsln}.  The degrees $n_\a$ of the 
polynomials continue to be given by (\ref{expon}).  The charges $Q_\a$ are
related to the values $q_\a(0)$ of the variables $q_\a$ at $\rho=0$ ({\it
i.e.}\ at spatial infinity) by \cite{lpsln}
\be
e^{q_\a(0)} = \prod_\beta (4 Q_\beta)^{4(M^{-1})_{\a\beta}}\ .\label{charge}
\ee

    In the case of $E_8$, the Toda equations (\ref{toda8}) 
are given by
\bea
&&q_1'' = e^{2q_1 - q_4}\ ,\qquad\qquad
q_2'' = e^{2q_2 - q_3}\ ,\nn\\
&& q_3'' = e^{-q_2 + 2 q_3 - q_4}\ ,\qquad
q_4'' = e^{-q_1 - q_3 + 2 q_4 - q_5}\ ,\nn\\
&&q_5'' = e^{-q_4 + 2 q_5 - q_6}\ ,\qquad
q_6'' = e^{-q_5 + 2 q_6 - q_7}\ ,\label{e8pbrane}\\
&&q_7'' = e^{-q_6 + 2 q_7 - q_8}\ ,\qquad
q_8'' = e^{-q_7 + 2 q_8}\ .\nn
\eea
 From (\ref{expon}), we find that the degrees $n_\a$ of the polynomials for
$e^{-q_\a}$ in this case will be $n_\a = 
\{ 136, 92, 182, 270, 220, 168, 114, 58\}$.  Solving for the 1248 
coefficients $a_{\a m}$ in terms of the eight independent parameters 
associated with the eight charges $Q_\a$ is mechanical, 
but somewhat involved.  Instead, we shall just present the results for
a subclass of solutions, where we truncate the system to the $D_4=O(4,4)$
subalgebra with simple roots $\{\vec a_{123},\vec b_{23},\vec b_{34}, 
\vec b_{45} \}$ (see Table 1).  This is a new solution that lies outside 
the $A_N$  solutions obtained in \cite{lpsln}.  For the $D_4$ 
truncation, we find
\bea
e^{-q_1} &=& c_1 c_3 - \ft18 c_3^6 -36 c_2 c_4 + 6 c_3^3 c_4 -
                 36 c_4^2 \nn\\
&&+ c_1 \rho + (3c_2 c_3 -\ft18 c_3^4) \rho^2 + c_2 \rho^3\nn\\
&& + \ft1{24} c_3^2 \rho^4 + \ft1{60} c_3 \rho^5 + \ft1{360} \rho^6\ ,\nn\\
e^{-q_3} &=& c_1 c_3 - 36 c_2^2 + 3 c_2 c_3^3 - \ft18 c_3^6 - 36c_2 c_4
            \nn\\
&& + 3 c_3^3 c_4 + ( c_1 - 3c_2 c_3^2 + 3 c_3^2 c_4) \rho\nn\\
&& + (3 c_3 c_4 - \ft18 c_3^4) \rho^2 + c_4 \rho^3 \nn\\
&& + \ft1{24} c_3^2 \rho^4 + \ft1{60} c_3 \rho^5 + \ft1{360} \rho^6\ ,\nn\\
e^{-q_5} &=& c_1 c_3 - 6 c_2 c_3^3 +\ft38 c_3^6 + 36 c_2 c_4 - 3c_3^3 c_4
\nn\\
&& +(c_1 - 6 c_2 c_3^2 + \ft12 c_3^5 - 3 c_3^2 c_4) \rho - (3c_2 c_3 \nn\\
&& - \ft38 c_3^4 + 3 c_3 c_4) \rho^2 - (c_2 + c_4 - \ft16 c_3^3) \rho^3\nn\\
&&+ \ft1{24} c_3^2 \rho^4 + \ft1{60} c_3 \rho^5 + \ft1{360} \rho^6\ ,
\label{d4tod}\\
e^{-q_4} &=& c_1^2 - 6 c_1 c_2 c_3^3 + \ft14 c_1 c_3^5 + \ft34 c_2c_3^7\nn\\
&&            -\ft1{32}c_3^{10} + 216c_2^2 c_3 c_4 - 45c_2 c_3^4 c_4 \nn\\
&&          +\ft32 c_3^7 c_4 + 216 c_2 c_3 c_4^2 - 9 c_3^4 c_4^2 \nn\\
&& +(\ft34 c_2 c_3^6 -\ft14 c_1 c_3^4 + 216c_2^2 c_4 - 36 c_2 c_3^3 c_4 \nn\\
&& + 216 c_2 c_4^2) \rho + (18 c_2^2 c_3^2 - \ft12 c_1 c_3^3 -
               \ft32 c_2 c_3^5\nn\\
&& + \ft3{32} c_3^8 + 18 c_2 c_3^2 c_4 - 3c_3^5 c_4 + 18 c_3^2 c_4^2)\rho^2
\nn\\
&& +(12c_2^2 c_3 -\ft12 c_1 c_3^2 - \ft12 c_2 c_3^4 +\ft1{24} c_3^7\nn\\
&& +12c_2 c_3 c_4 - 2c_3^4 c_4 + 12 c_3 c_4^2)\rho^3 + (3c_2^2 \nn\\
&&+ 3 c_4^2-\ft14 c_1c_3 +\ft14c_2 c_3^3 + 3c_2 c_4 -\ft12c_3^3 c_4)
\rho^4\nn\\
&& +(-\ft1{20}c_1 + \ft3{20}c_2 c_3^2 +\ft1{240}c_3^5)\rho^5 + 
\ft{7}{720}c_3^4\rho^6\nn\\
&& +\ft1{180}c_3^3 \rho^7 +\ft1{480}c_3^2 \rho^8 + \ft1{2160} c_3 \rho^9
+\ft1{21600}\rho^{10}\ ,\nn
\eea
where $c_i$ ($i=1,2,3,4$) are arbitrary constants. 

    The relation between the four arbitrary constants in (\ref{d4tod}) and
the four independent charges in the $D_4$ Toda extremal $p$-brane solutions
is given by (\ref{charge}).  The mass of the solution is expressible in
terms of the charges {\it via} a polynomial equation whose degree is equal
to the dimension of the Weyl group of the Lie algebra characterising the Toda
equations \cite{lpsln}.  In the case of the $D_4$ example above, this means
that there will be a polynomial of degree 192 relating the mass and the
charges.  Obtaining the relation between the mass and the eight charges of 
the $E_8$ Toda $p$-branes would require a more challenging calculation, since
the polynomial is of degree 696,729,600.

\end{document}